\documentclass[12pt]{aastex}

\usepackage{natbib}
\usepackage{graphicx}

\usepackage{amssymb}

\def\bnabla{{\mathbf{\nabla}}}
\def\bOmega{{\mathbf{\Omega}}}

\def\bJ{{\mathbf{J}}}
\def\bv{{\mathbf{v}}}
\def\bE{{\mathbf{E}}}
\def\bF{{\mathbf{F}}}
\def\be{{\mathbf{\hat e}}}

\def\btimes{{\mathbf{\times}}}
\def\bR{{\mathbf{R}}}
\def\bB{{\mathbf{B}}}
\def\m{\,{\rm m}} 
\def\s{\,{\rm s}}

\def\refnew#1{(\ref{#1})}

\begin{document}
 
  \title{Spontaneous Axisymmetry Breaking of Saturn's External Magnetic Field}  
\author{Peter Goldreich}
\affil{Institute for Advanced Study, Princeton, New Jersey, USA; pmg@ias.edu}
\author{Alison J. Farmer}
\affil{Department of Astronomy, Harvard University, MS-51, 60 Garden St., Cambridge, Massachusetts, USA; afarmer@cfa.harvard.edu}

\begin{abstract}
Saturn's magnetic field is remarkably axisymmetric. Its dipole axis is
inclined by less than $0.2^\circ$ with respect to its rotation
axis. Early evidence for nonaxisymmetry came from the periodicity of
Saturn's kilometric radio bursts (SKR). Subsequently, percent
variations of the SKR period were found to occur on timescales of
years. A recent breakthrough has been the direct detection of a
nonaxisymmetric component of the field that rotates with a period
close to that of the SKR.  Because this component's magnitude varies
only weakly with distance from Saturn, it must be supported by
currents external to the planet. These currents flow along field lines
that connect plasma in the equatorial region of the outer
magnetosphere (the plasma disk) to the high latitude ionosphere. The
plasma originates from mass lost by the planet's rings and moons.  It
is tightly coupled to the magnetic field and its motion can be
ascribed to the large scale interchange of flux tubes.  Heavily loaded
tubes drift outward and are replaced by lightly loaded ones which
drift inward. {\it This process of rotationally driven convection is
responsible for breaking the axisymmetry of Saturn's external magnetic
field.} The convection pattern rotates rigidly with the angular
velocity of the plasma at its source. Its rotation provides the clock
that controls the periods of both the SKR bursts and the
nonaxisymmetric magnetic perturbations. We distinguish two types of
currents. Those that flow in the azimuthal direction in both the
ionosphere and the plasma disk provide a radial force that enables the
plasma to corotate with the planet's spin.  They are responsible for
stretching magnetic field lines and thus exposing the polar ionosphere
to the incoming solar wind. Other currents flow in the latitudinal
direction in the ionosphere and the radial direction in the plasma
disk. These act to transfer angular momentum from the planet's spin to
the outflowing plasma. They slow down the rotation rate of the
ionosphere below that of the underlying atmosphere and are the reason
the clock referred to above runs slow. Moreover, these currents are
the source of nonaxisymmetric magnetic perturbations whose strength
varies inversely with radial distance in the planet's equatorial
plane. Quantitative agreement with the magnitude of these
perturbations requires a mass loss rate of order $10^4{\, \rm g\,
s^{-1}}$, similar to that believed to come from Saturn's E-ring.
\end{abstract}
 
\section{Introduction} 
\label{sec:intro}

Saturn's atmosphere exhibits strong ($\lesssim 400 \m \s^{-1}$) and
stable (over decadal time scales) zonal winds.  This precludes
assigning a unique rotation period to its outer layers. Although the
planet is fluid throughout, its deep interior must be in near solid
body rotation \citep{liu06}. Thus the internal rotation rate might be
revealed by observing the nonaxisymmetric components of the planet's
magnetic field.\footnote{This technique applied to Jupiter has found a
rotation period constant to within seconds over 50 years.}  However,
application of this technique to Saturn has been hampered by the
extreme axisymmetry of the planet's magnetic field.  

\emph{Voyager} observations of Saturn Kilometric Radiation (SKR)
bursts coming from the planet's auroral regions showed a periodicity
of 10h 39 min 24 $\pm$ 7 s \citep{des81}. The periodicity was
suspected to arise from a small non-axisymmetry of the planet's
internally generated magnetic field. Detections of small in situ
magnetic anomalies from \emph{Voyager} and \emph{Pioneer 11}
magnetometers were also reported \citep{gia04}, with periods
consistent with that of the SKR. \emph{Ulysses} observations of the
variability of the SKR period, by of order 1 \% on timescales of 1
year from 1994 to 1997 \citep{gal00}, challenged its interpretation as
the rotation period of the planet's deep interior.\footnote{A change
of 1 \% in spin period of the planet's core over such a short
timescale is energetically impossible.} \emph {Cassini} confirmed this
variability, measuring an SKR period of 10h 45min 45 $\pm$45s on
approach to Saturn in 2004 \citep{gur05}.  Most recently, magnetometer
data obtained by \emph{Cassini} showed a small ($ \sim$ few nT) signal
with period 10h 47min 6 $\pm$40 s \citep{gia06} that was stable during
14 months of observation. \citet{gia06} suggested that this period
might be that of Saturn's interior spin. However, the decline of the
perturbation amplitude with radius is too slow to be due to a current
source within Saturn, so this interpretation cannot be
correct. Instead, we propose that the non-axisymmetric component of
Saturn's external magnetic field is generated by rotationally driven
convection in the planet's magnetosphere. The convection transports
plasma from the inner magnetosphere to the magnetopause where it joins
the solar wind. The plan of the paper is as follows. In \S
\ref{sec:electro} we provide a simplified version of equations
governing rotationally driven magnetospheric convection. We apply
these equations in \S \ref{sec:mag} to estimate the nonaxisymmetric
magnetic perturbations it produces. \S \ref{sec:clock} is devoted to a
discussion of the clock that controls the perturbations' rotation rate.
A short summary is given in \S \ref{sec:sum}.

\subsection{Nominal parameters}
\label{subsec:parameters}

We adopt cgs units for length, mass, time, and Gaussian units for
electrodynamical quanitites.  In order to focus our discussion on
Saturn, we provide numerical estimates along with some of the major
equations. The parameters used in these evaluations are displayed in
Table \ref{tab:nom}.
\begin{table}
\centering
\begin{tabular}{clc}
\hline
Quantity& Value adopted&  Notes\\
\hline
$R$&$6 \times 10^9$ cm&\\
$\Omega$&$1.6 \times 10^{-4} \rm{~s}^{-1}$&\\
$B_P$&0.2 G&\\
$\Sigma_p$&$10^{13} \; \Sigma_{p,13} \,\rm{~cm~s}^{-1}$& [1]\\
\hline
$a_i$&$ 4R \; a_{i,4R}$&[2]\\
$\Delta \phi_i$&$\pi \; \Delta\phi_{i,\pi}$&\\
$\dot{M}$&$10^4 \; \dot{M}_4 \, \rm{~g~s}^{-1}$&[3]\\
\hline
$\mu$&$3 \times 10^{-24}$ g&\\
$\gamma$&1.4&\\
$T$&$140 \; T_{140} \,$ K&[4]\\
$\nu$&$10^8 \; \nu_8 \rm{~cm}^2\rm{~s}^{-1}$&[5]\\
$\rho^I$&$3 \times 10^{-13} \; \rho^I_{-12.5} \rm{~g~cm}^{-3}$&[6]\\
$g$&$1 \times 10^3 \rm{~cm~s}^{-2}$&\\
\hline
\end{tabular}
\caption{The numerical quantities adopted in order of magnitude
calculations throughout the paper.  The meanings of the symbols are
given in the text where first used. Key to numbered notes: [1] scaled
between estimates for low and high
latitude ionosphere \citep{atreya84}; [2] scaled to orbit of E-ring; [3]
scaled to estimates for plasma production from
Saturn's E-ring \citep{leisner06}; [4] scaled to atmospheric temperature
at base of ionosphere \citep{atreya84}; [5] scaled to eddy diffusion
coefficient at homopause \citep{atreya84}; [6] scaled between estimates
for low latitude ionosphere
and auroral ionosphere \citep{atreya84}.}
\label{tab:nom}
\end{table}

\section{Basic Electrodynamics}
\label{sec:electro}

\subsection{Rotationally driven magnetospheric convection}
\label{subsec:past}

The topic has an extensive history, but this is not the place to
review it.  Instead, we point to a few influential papers that aided
our understanding.  We have done nothing more than to apply what
we learned from reading the literature.

Equations governing rotationally driven convection were formulated by
\citet{chen77} and by \citet{hill81}. Many applications have been made
to the outward transport of mass from the Io plasma
torus. \citet{pontius89} includes a clear discussion of different
approaches to this problem.

Progress in solving the equations referred to above has been slow. In
retrospect, this is not surprising. They are nonlinear set, which probably
precludes finding analytic solutions. Moreover, realistic applications
are faced with including a continuous supply of plasma along with boundary
conditions that simulate its loss to the solar wind at the magnetopause. 

Perhaps the most ambitious attempt at a realistic solution is that by
\citet{yang94} who applied the Rice convection model to the Io torus.
They investigated an initial value problem, the instability of a torus
of finite width. An active source of plasma was not included. Long
fingers were found to grow radially outward from the torus. This is
not surprising. The initial state is analogous to that of a heavier
fluid resting on top of a lighter one which is Rayleigh-Taylor
unstable. But unlike the standard Rayleigh-Taylor instability which
takes place for constant gravitational acceleration, the instability
of the plasma torus is driven by centrifugal acceleration which
increases linearly outward. This increase allows narrow fingers to
run away from the more slowly developing, thicker modes that might
otherwise subsume them.

In the absence of anything better, we adopt a simplified picture of
steady-state magnetospheric convection. A tongue of plasma flows
outward from a torus of neutral material that is escaping from
Enceladus. Plasma is continuously created inside the torus by
ionization of this material. It exits in a tongue which flows
outward. Except within the tongue, flux tubes outside the torus drift
inward due to ``fringing'' electric fields surrounding the tongue region. As they cross the torus, the inward-moving tubes are loaded with freshly created
plasma. Then their trajectories bend around so that they join the back
of the tongue. In this manner, plasma is continuously removed from the
entire torus even though the tongue emanates from only a limited range
of azimuth. 

In order for the convection pattern to remain steady, plasma must consistently outflow from the same range of azimuth in the rotating frame. Outflow occurs from the densest part of the torus, and so if a single tongue is to carry the outgoing material, its base must always be refilled fast enough so that no other longitudes in the torus can accumulate more plasma. Inwardly drifting tubes, which empty the rest of the torus of plasma, drift more slowly than those moving outwards, because the electric fields in the inward-drifing tubes cannot be larger than the internal tongue fields which they ``fringe''. Only if the base of the tongue spans $\Delta \phi
_i \gtrsim \pi$ radians can the inwards drift velocity be large enough to keep the plasma content in the rest of the torus lower than in the tongue region.

We have been able to bolster this description with
simplified models of electrostatic fields, but much remains to be done
before anything rigorous might emerge. A serious technical issue is
that a smooth tongue of plasma is likely to
develop narrower fingers as discussed above. Fortunately, the
conclusions of our investigation are insensitive to this
possibility. However, it would certainly impede a rigorous calculation
of the convection pattern. That will have to be left for the future.

\subsection{Notation}

We adopt spherical polar coordinates $r,\, \theta,\, \phi$ and work in
the inertial frame. Saturn's magnetic field is approximated as a spin
aligned dipole.  The ionosphere is taken to rotate with uniform
angular velocity $\Omega$.\footnote{Except in \S \ref{sec:clock} where
$\Omega$ denotes the angular velocity of the deep atmosphere.} Where
necessary, superscripts $^M$ and $^I$ are used to distinguish
magnetospheric and ionospheric quantities, and $^{Ip}$ to denote the
direct (Pedersen) component of the ionospheric current. $B_r$ and
$B_\theta$ are the components of the unperturbed magnetic field at a
general field point. $R$ is Saturn's radius and $a$ is the orbital
radius in the equatorial plane. $B_P$ is the magnetic field intensity
on the equator in Saturn's ionosphere and $B_z$ is the component of
the {\it vertical} magnetic field in the magnetic equator at $a$;
\begin{equation}
B_z=-\left(R\over a\right)^3 B_P\, .
\label{eq:BzBP}
\end{equation}
Height integrated current densities and electrical conductivities are
indicated by $J$ and $\Sigma$, respectively. The surface mass density
in the magnetosphere is denoted by $\sigma$.

\subsection{Dipole magnetic fields}

Components of a spin aligned dipole magnetic field take the form
\begin{equation}
B_r=\frac{2M\cos\theta}{r^3}\, \quad\quad\quad B_\theta=\frac{M\sin\theta}{r^3}\, ,
\label{eq:compB}
\end{equation}
where $M$ is the dipole moment. The field magnitude is
\begin{equation}
B=\left(B_r^2+B_\theta^2\right)^{1/2}=\frac{M}{r^3}(1+3\cos^2\theta)^{1/2}\, .
\label{eq:magB}
\end{equation}
\noindent A individual field line is labeled by either the colatitude of its
footprint at $r=R$, denoted by $\theta_0$, or by its maximum radial extent
$a$; $a\sin^2\theta_0=R$. Its shape is described by:
\begin{equation}
r\sin^2\theta_0 = R\sin^2\theta\, \quad\quad {\rm or}\quad\quad r=a\sin^2\theta\, .
\label{eq:lines}
\end{equation}

We are interested in field lines that connect to the planet at high
latitudes. Thus we simplify our expressions by setting
$\sin\theta_0=(R/a)^{1/2}$ and $\cos\theta_0=1$.

\subsection{Magnetospheric currents}

Consider an element of cold plasma that is nearly corotating with and
slowly drifing away from the planet. Centrifugal balance and angular
momentum conservation require that the height integrated current densities
which pass through the element satisfy\footnote{We neglect the planet's
gravity in equation \refnew{eq:JphiM}}
\begin{equation}
\frac{J_\phi^M B_z}{c}=-\sigma\Omega^2 a\, 
\label{eq:JphiM}
\end{equation}
and
\begin{equation}
\frac{J_a^M B_z}{c}=-2\sigma\Omega{\dot a}\, 
\label{eq:JaM}
\end{equation}
respectively. 

The ratio
\begin{equation}
\frac{J_a^M}{J_\phi^M}=\frac{2 \dot a}{\Omega a}\, 
\label{eq:ratioJM}
\end{equation}
is small in the inner magnetosphere but becomes of order unity in its
outer regions.

\subsection{Ionospheric direct currents}

The horizontal divergence of the magnetospheric currents, $\bnabla_{\rm
2d}\cdot\bJ^M$, flows along magnetic field lines and closes in the
ionosphere. This determines the components of the direct (Pedersen)
current in the ionosphere. They read
\begin{equation}
J_\phi^{Ip}=\frac{1}{R}\frac{da}{d\theta_0}\frac{J_\phi^M}{2}=
-\frac{c\sigma\Omega^2 R}{B_P}\left(a\over R\right)^{11/2}\, ,
\label{eq:JphiI}
\end{equation}
and
\begin{equation}
J_\theta^{Ip}=\frac{a}{R\sin\theta_0}\frac{J_a^M}{2}=\frac{c\sigma\Omega
R}{B_P}\left(a\over R\right)^{11/2}\frac{\dot a}{a}\, .
\label{eq:JthetaI}
\end{equation}
Their ratio is given by
\begin{equation}
\frac{J_\theta^{Ip}}{J_\phi^{Ip}}=-\frac{\dot a}{\Omega a}\, .
\label{eq:ratioJI}
\end{equation}

The ionospheric Hall current, $\bJ^{Ih}$, is not fixed in this
manner. Because its horizontal divergence
vanishes,\footnote{Provided the Hall conductivity is independent of
position as we assume it to be.} its determination requires
knowledge of the ionospheric electric field.

\subsection{Ionospheric electric field}

The height-integrated current density, $\bJ^I$, and electric field,
$\bE^I$, are related by\footnote{By setting $\sin\theta_0=(R/a)^{1/2}$
and $\cos\theta_0=1$, we neglect both $B^I_\theta$ and the effects of
the parallel conductivity.}
\begin{equation}
\bJ^I=\bJ^{Ip}+\bJ^{Ih}=\Sigma_p\bF^I+\Sigma_h({\hat {\bf b}}\times \bE^I)\, ,
\label{eq:ohm}
\end{equation}
where $\bF^I\equiv \bE^I+(\bOmega\times\bR)\times \bB^I/c$ is the
Lorentz force, $\Sigma_p$ and $\Sigma_h$ are the height-integrated
Pederson and Hall conductivities, and ${\bf b}\equiv
\bB^I/|\bB^I|$. Taking the divergence of equation \refnew{eq:ohm}
yields
\begin{equation}
\bnabla_{2d}\cdot \bJ^I=\bnabla_{2d}\cdot
\bJ^{Ip}=\Sigma_p\bnabla_{2d}\cdot\bE^I\, .
\label{eq:div2dJI}
\end{equation}
$\bnabla_{2d}\cdot\bJ^{Ih}=0$ because $\bnabla_{2d}\times\bE^I=0$;
$\bE^I$ is a potential field. A full solution for $\bE^I$ would
involve setting $\bE=-\bnabla\Phi^I$ and then solving Poisson's
equation $\nabla^2\Phi^I=-\Sigma_p\bnabla_{2d}\cdot\bJ^{Ip}$. As
detailed in \S \ref{subsec:past}, for realistic conditions, this has
proven to be a difficult task. Fortunately, a simpler procedure
suffices for the purposes of the current investigation.

We set 
\begin{equation}
F_\theta = \frac{J_\theta^{Ip}}{\Sigma_p}\, \quad\quad {\rm and}
\quad\quad E_\phi = \frac{J_\phi^{Ip}}{\Sigma_p} \, ,
\label{eq:EIc}
\end{equation}
with
\begin{equation}
F_\theta\equiv E_\theta-\frac{2\Omega R\sin\theta}{c}B_P\, .
\label{eq:LorFth}
\end{equation}

This procedure does a good job evaluating $\bE^I$ in the portion of
the ionosphere that is magnetically connected to the outgoing tongue
of magnetospheric plasma. However, it does not permit a determination
of the fields that fringe this region. These control the
inward flow of depleted plasma tubes.

\subsection{$\bE\btimes\bB$ drift}

Just above the ionosphere, the plasma drifts at velocity
\begin{equation}
\bv=c\left(\frac{E_\phi}{2B_P}\be_\theta-\frac{E_\theta}{2B_P}\be_\phi\right)\, .
\label{eq:vdrift}
\end{equation}
Projecting down to the magnetosphere, we obtain
\begin{equation}
\frac{\dot a}{a}=-\frac{2\dot\theta_0}{\theta_0}=-\frac{cE_\phi}{R B_P}\left(a\over
R\right)^{1/2}=\frac{c^2\sigma\Omega^2}{\Sigma_p B_P^2}\left(a\over R\right)^6 \,.
\label{eq:adot}
\end{equation}
and 
\begin{equation}
\frac{\Delta\Omega}{\Omega}=\frac{cF_\theta}{2\Omega R B_P}\left(a\over
R\right)^{1/2}=\frac{c^2\sigma}{2\Sigma_p B_P^2}\left(a\over
R\right)^6 \frac{\dot a}{a}\, .
\label{eq:OmDelOm}
\end{equation}
Here $\Delta\Omega$ is the angular velocity at which the plasma in the
tongue slips relative to the rotation at the top of the ionosphere.

For future reference, we note that 
\begin{equation}
\left(\frac{\dot a}{a}\right)^2=2\Omega\Delta\Omega\, .  
\label{eq:DelOm} 
\end{equation}

\subsection{Coupling to rate of mass loss}

Suppose the tongue covers $\Delta\phi(a)$ in azimuth where
$\Delta\phi$ is a function of $a$ to be determined later.
Then
\begin{equation}
{\dot M}=\Delta\phi a{\dot a}\sigma \, .
\label{eq:Mdot}
\end{equation}
Substituting for ${\dot a}$ using equation \refnew{eq:adot}, we
arrive at 
\begin{equation}
{\dot M}=\frac{\Delta\phi
c^2\sigma^2\Omega^2 R^2}{\Sigma_p B_P^2}\left(a\over R\right)^8\, ,
\label{eq:dotM}
\end{equation}
from which we obtain
\begin{equation}
\sigma=\frac{|B_P|}{c\Omega R}\left(\frac{\Sigma_p\dot M}{
\Delta\phi}\right)^{1/2}\left(R\over a\right)^4\, .
\label{eq:sigma}
\end{equation}
Next we replace $\sigma$ in ${\dot a}/a$ which yields
\begin{equation}
\frac{\dot a}{a}=\frac{\Omega c}{ R |B_P|}\left(\frac{ \dot
M}{\Sigma_p\Delta\phi}\right)^{1/2}\left(a\over R\right)^2\, .
\label{eq:adotM}
\end{equation}

\subsection{Steady-state scalings}

Flux freezing implies that $\sigma/B_z$ is independent of $a$ in a
sourceless, steady-state flow. Consequently, $\Delta\phi$ at $a$ is
related to its initial value at $a_i$ by
\begin{equation}
\Delta\phi=\left(\frac{a_i}{a}\right)^2\Delta\phi_i\, .
\label{eq:DeltaPhia}
\end{equation}
Thus
\begin{equation}
\sigma=\frac{|B_P|}{c\Omega a_i}\left(\frac{\Sigma_p\dot M}{
\Delta\phi_i}\right)^{1/2}\left(R\over a\right)^3\, .
\label{eq:sigmap}
\end{equation}
\begin{equation}
\dot a=\frac{R\Omega c}{a_i |B_P|}\left(\frac{ \dot
M}{\Sigma_p\Delta\phi_i}\right)^{1/2}\left(a\over R\right)^4\, .
\label{eq:adotap}
\end{equation}

Somewhat arbitrarily, we adopt $a_o$, the value of $a$ where ${\dot
a}=\Omega a$, or equivalently where $\Delta\Omega=-\Omega/2$, as the
outer radius of the region in which partial corotation applies;
\begin{equation}
\left(a_o\over R\right)=\left(\frac{a_i|B_P|}{c}\right)^{1/3}
\left(\Sigma_p\Delta\phi_i\over {\dot M}\right)^{1/6} \sim 21
\left(\frac{a_{i,4R}^2 \Delta \phi_{i,\pi} \Sigma_{p,13}}{\dot
M_4}\right)^{1/6}\, .
\label{eq:ao}
\end{equation}

The timescale $a/\dot a$ for outward plasma transport decreases with distance from the planet. Plasma near the source at $a_i$ doubles its radial distance in a time $\sim 10$ days. The constancy of the observed magnetic period on far longer timescales implies that the convection pattern remains steady for many dynamical timescales of the source region.

\section{Magnetic Perturbations}
\label{sec:mag}

\subsection{Opening of the polar cusp}

The azimuthal current $J^M_\phi$ creates a radial component of
magnetic field just above and below the tongue of outgoing plasma;
\begin{equation}
\Delta B_r=\frac{2\pi J^M_\phi}{c}{\rm sgn}(z)\, . 
\label{eq:DeltaBr}
\end{equation}
Applying equations \refnew{eq:sigmap} and \refnew{eq:ao}, we find that
these field lines bulge outward relative to vertical by an angle
\begin{equation}
\tan\alpha=\frac{|\Delta B_r|}{B_z}=\frac{2\pi\sigma\Omega^2 a}{B_z^2}
=\frac{2\pi\Omega a_o\Sigma_p}{c^2}\left(a\over a_o\right)^4 \sim 1.4
\left(\frac{a_{i,4R}^2 \Delta \phi_{i,\pi} \Sigma_{p,13}^7}{\dot
M_4}\right)^{1/6}\, .
\label{eq:alpha}
\end{equation}
The stretching of these field lines is a consequence of the centripetal
acceleration they impart to the plasma.

If $\alpha$ attains a substantial value at $a_o$, the exposure of
Saturn's polar cap to the incoming solar wind will be enhanced along
the range of longitudes subtended by the outer parts of the
tongue.\footnote{The field lines bulge only above and below the outgoing
plasma. Because $\Delta \phi \propto a^{-2}$, $\Delta \phi_0$ may be
quite small, even if $\Delta \phi_i$ is of order $\pi$.}  This could
account for the preferential emission of the SKR within a narrow range
of longitudes.

\subsection{Magnetic anomalies measurable by spacecraft}

The radial current in the tongue of outgoing material at $a$ is given by
\begin{equation}
I^M_a=a\Delta\phi J^M_a=-\frac{2c\Omega{\dot M}}{B_z}\, .
\label{eq:IMa}  
\end{equation}
Thus 
\begin{equation}
\frac{dI^M_a}{da}da=-\frac{6c\Omega {\dot M}}{aB_z}da\, 
\label{eq:dI/da}
\end{equation}
is the current that flows along the strip of field lines connecting
the ionosphere (in each hemisphere) to the tongue between $a$ and
$a+da$. It is notable that $I_a^M$ depends on the rate of mass loss
but not on the pattern of outflow.

To estimate the magnetic perturbation produced by this current, we
approximate it as flowing along a thin wire of infinite extent which
passes the observation point at distance $s$.  Then
\begin{equation}
\frac{d\delta B}{da}\approx \frac{2}{c s}\frac{dI^M_a}{da}\approx
\frac{12\Omega{\dot M}}{RB_P s}\left(a\over R\right)^2\, .
\label{eq:ddelBda}
\end{equation}
We focus on perturbations in the equatorial region of the inner
magnetosphere, since this is where the \emph{Cassini} magnetometer has
found evidence for non-axisymmetric field components. For an observer
in the equatorial plane at radius $\varpi$, the minimum distance $s$
between the observer and a ``wire'' of dipole field-aligned current
which passes through the equator at $a \gg \varpi$ is $s \simeq
\varpi$.  Substituting for $s$ in Eq. \ref{eq:ddelBda} and integrating
to $a_o$, we find
\begin{equation}
\delta B\approx \frac{4\Omega{\dot M}}{B_P\varpi}\left(a_o\over
R\right)^3\approx \frac{4\, a_i\Omega}{c\varpi}\left(\Sigma_p{\dot
M}\Delta\phi_i\right)^{1/2} \sim 1.2 \times 10^{-5} {\rm~G~}
(\Delta\phi_{i,\pi} \dot M_4 \Sigma_{p,13})^{1/2} \, .
\label{eq:delB}
\end{equation}

The above estimate for the magnitude of the magnetic perturbations
ignores the contribution from the return current that comes from
$a>a_o$ and closes the circuit. Unfortunately, not much can be deduced
about the geometry of the return current since the field lines it
flows along are likely to be strongly perturbed by the solar
wind. However, it is possible that it could act to reduce the
perturbation magnitude, perhaps by a factor 2. We have also ignored
the contribution from the currents that flow in the opposite (N versus
S) hemisphere. In the case of perfect N-S symmetry, magnetic
perturbations from the two hemispheres would exactly cancel in the middle of the tongue but not elsewhere. In reality, at most times
\footnote{The exception being near times of equinox.} this cancelation
is likely to be small, since the ionospheric conductivity will differ
in the northern and southern polar ionospheres. Near solstice, almost
all the current will flow through the summer hemisphere. At these times
the estimate for $\delta B$ should be doubled.

\begin{figure}[h]
\centering
\includegraphics[width=6.5in]{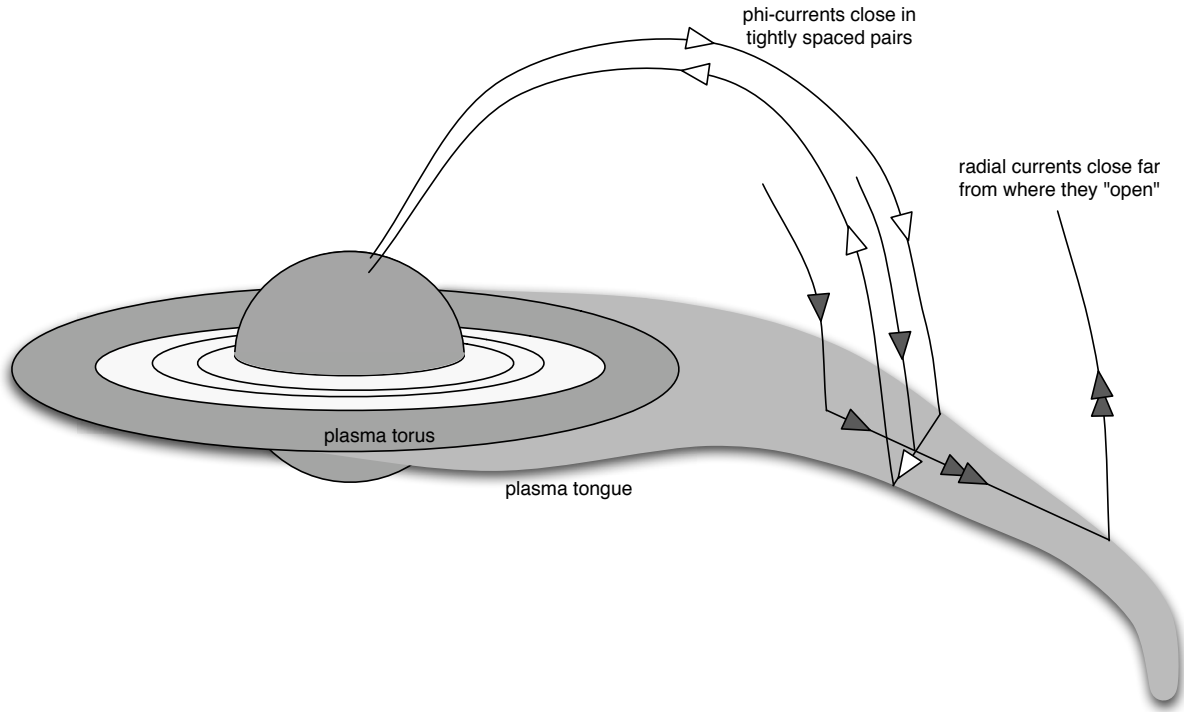}
\caption{The currents associated with the plasma tongue in Saturn's
magnetosphere: azimuthal currents occur in close opposite pairs, and
so their contribution to magnetic perturbations is small. }
\label{fig:tongue}
\end{figure}

 As illustrated in Figure \ref{fig:tongue}, the field-aligned
(Birkeland) currents supplying $J^M_\phi$ to the tongue occur in
closely spaced, oppositely directed pairs. The magnetic perturbations
produced by these ``line dipoles'' are small at distances large
compared to their spacing. Thus, although they are typically smaller,
the field-aligned currents feeding $J^M_a$ dominate the magnetic
perturbations present in the magnetosphere.

The fragmentation of the plasma tongue into narrower structures would
have a minor effect on the magnetic perturbations produced by the
currents supplying $J_a^M$, but would further weaken those produced by the
currents supplying $J_\phi^M$.

Analysis of the variation of the measured components of $\delta B$ with
position in the magnetosphere would test our model.

\section{The Clock}
\label{sec:clock}

As described in \S \ref{sec:intro}, a somewhat slow and imperfect
clock controls the quasi-periodic behavior of SKR bursts. In our
scenario the same clock controls the magnetic anomalies. Both
\emph{Voyager}- and \emph{Cassini}-era measurements of SKR and
magnetic periods are consistent with this picture.

\subsection{Where is the clock located?}

The clock is located in the inner magnetosphere and beats at the
period of rotation of the asymmetry associated with the inner portion
of the plasma outflow.\footnote{Our working hypothesis is that the
E-ring centered at $a\approx 4R$ around the orbit of Enceladus is the
dominant plasma source.} As explained below, the clock's period
propagates throughout the magnetosphere provided the differential
rotation of the latter remains time invariant.

Suppose that a source of material located at radius $a_i$ feeds a
plasma torus rotating at $\Omega_i$. The torus is unstable and sends
out a ``tongue'' of plasma centered on a fixed azimuth $\phi_c(a_i)$
in the rotating frame. Plasma at the center of the tongue moves
outwards at radial speed $\dot a_c(a)$, and orbits at angular speed
$\dot \phi_c(a)$, where $\dot \phi_c(a)\lesssim \Omega_i$. In steady-state,
the shape of the tongue's centerline would be determined by
\begin{equation}
\phi_c(a)=\phi_c(z_i)+\int_{a_i}^a da^\prime\,\frac{\dot
\phi_c(a^\prime)}{\dot a_c(a^\prime)}\, .
\label{eq:ctongue}
\end{equation}

Viewed from a nonrotating frame, the tongue is a steady structure
rotating at pattern speed $\Omega_i$.  At radius $a$ and time $t$,
the apparent azimuth of the tongue's centerline is given by
\begin{equation}
\phi_{\rm obs} = \phi_c(a)+\Omega_i t,
\end{equation}
and so $\dot \phi_{\rm obs} = \Omega_i$, regardless of the run of
differential rotation across the magnetosphere.

\subsection{Why is the clock slow?}

SKR and magnetic periods determined by Cassini are longer than those
associated with the motion of any atmospheric features. The most
plausible explanation is that the ionosphere rotates more slowly than
the atmosphere below it because magnetic torques are transferring
angular momentum from it to the plasma tongue.\footnote{\citet{huang89}
concluded that the rotation of Jupiter's ionosphere is slowed in this
manner.} 

Since the magnetic torque increases sharply with increasing latitude,
this is also a plausible explanation for the observed decline of the
magnetosphere's angular velocity with increasing distance from
Saturn. We shall show that this scenario is qualitatively reasonable,
in contrast to models based on the slippage of the rotation of the
magnetospheric plasma relative to the rotation of the ionosphere.
Most of the way out, the plasma maintains good corotation with the
part of the ionosphere to which it is connected, but the ionosphere is
subcorotating with respect to the underlying atmosphere. The observed
clock frequency is the rotation rate of the ionosphere where it
connects to the inner part of the plasma tongue.

\subsubsection{Steady-state rotation of the ionosphere}

We analyze a simple model for the steady-state rotation of the
ionosphere and underlying atmosphere.  It assumes axial symmetry and
considers only vertical transport of angular momentum. Deep
atmospheric layers are taken to rotate rigidly with angular velocity
$\Omega$.  We work within the approximation of an isothermal
atmosphere with sound speed $c_s$, scale height $H=c_s^2/\gamma g$,
buoyancy frequency $N^2=(\gamma-1)g/(\gamma H)$, and eddy diffusivity
$\nu$. The ionosphere is taken to be a single layer rotating at
the angular velocity $\Omega^M$ of the part of the plasma tongue to
which it is magnetically connected. 

We modify equation \refnew{eq:JaM} to allow for the nonuniform rotation
rate, $\Omega^M$, of the magnetospheric plasma. The torque per unit
$a$ applied to the tongue of outgoing plasma reads
\begin{equation}
\frac{dT_B^M}{da}={\dot M}\frac{d}{da}\left(\Omega^Ma^2\right)\, .
\label{eq:TorM}
\end{equation}
Thus the magnetic torque per unit $\theta_0$ on the northern
ionosphere is given by
\begin{equation}
\frac{dT_B^I}{d\theta_0}=\frac{{\dot
M}R^2}{2}\frac{d}{d\theta_0}\left(\Omega^M\over\theta_0^4\right)\, .
\label{eq:dTI}
\end{equation}
In steady state, the torque must be constant with depth below the
ionosphere. Provided the torque is not too large, a stable,
steady-state, shear flow is established in which the viscous torque
given by
\begin{equation}
\frac{dT_\nu}{d\theta_0}=-2\pi R^4\theta_0^3\rho\nu\frac{d\Omega^A}{dz}\, ,
\label{eq:dTnu}
\end{equation}
carries angular momentum up from the atmosphere to the
ionosphere. Equating the viscous torque to the magnetic torque yields
\begin{equation}
\frac{d\Omega^A}{dz}=\frac{{\dot M}}{4\pi
R^2\rho\nu\theta_0^3}\frac{d}{d\theta_0}\left(\frac{\Omega^M}{\theta_0^4}\right)
\, .
\label{eq:dlnOmdt}
\end{equation}
The shear flow is stable where the Richardson criterion is satisfied,
that is where equation (\ref{eq:dlnOmdt}) predicts 
\begin{equation}
R\theta_0\frac{d\Omega}{dz} \lesssim N\, .
\label{eq:critshear}
\end{equation}
In the stable regime, 
\begin{equation}
\Omega-\Omega^M\approx -\frac{{\dot M}H}{4\pi R^2\rho^I\nu
\theta_0^3}\frac{d}{d\theta_0}\left(\frac{\Omega^M}{\theta_0^4}\right)
\approx \frac{{\dot M}\Omega H}{\pi R^2\rho^I\nu\theta_0^8}\, .
\label{eq:delOmstable}
\end{equation}
Here we have set $\Omega^M=\Omega$ in the final step. This is a good
approximation since $\Omega-\Omega^M\ll\Omega$ in the stable regime.

The boundary of the stable regime occurs where the stability criterion
is violated just below the ionosphere. We use the symbol
$\Theta_0$ to denote the value of $\theta_0$ at this boundary. At
$\Theta_0$,
\begin{equation}
\frac{\dot M}{4\pi R\rho^I\nu
N\Theta_0^2}\frac{d}{d\theta_0}\left(\Omega^M\over
\theta_0^4\right)\approx -1\, .
\label{eq:Theta}
\end{equation}
An approximate solution for $\Theta_0$, and the corresponding $a_{\rm
crit} /R\equiv \Theta_0^{-2}$, follows from setting $\Omega^M=\Omega$:
In this manner we arrive at
\begin{equation}
\frac{a_{\rm crit}}{R}=\left(1\over \Theta_0^2\right)\approx
\left(\frac{\pi R\rho^I\nu N}{\dot M \Omega}\right)^{2/7} \sim 9.1
\left(\frac{\nu_8 \rho^I_{-12.5}}{\dot M_4
T_{140}^{1/2}}\right)^{2/7}\, .
\label{eq:Theta0}
\end{equation}

The unstable layer penetrates deeper into the atmosphere poleward of
$\Theta_0$. Because the angular velocity gradient in the stable layer
is inversely proportional to density, and the density increases
exponentially with depth, the thickness of the unstable layer
increases logarithmically with decreasing $\theta_0$. An estimate
for $\Omega-\Omega^M$ is obtained by multiplying the critical angular
velocity gradient from equation \refnew{eq:critshear} by the thickness
of the unstable layer. The following
expression provides a good fit to $\Omega-\Omega^M$ for all values of
$\theta_0$ or $a/R$. We express it in terms of the latter for ease of
comparison with data on the rotation of plasma in Saturn's
magnetosphere.
\begin{equation}
1-\frac{\Omega^M}{\Omega} \approx \frac{NH}{\Omega R}\left(a\over
R\right)^{1/2} \ln\left[1+\left(a\over
a_{\rm crit}\right)^{7/2}\left(\frac{\Omega^M}{\Omega}+\frac{a}{2\Omega}
\frac{d\Omega^M}{da}\right)\right]\, .
\label{eq:DelOmOmsImp}
\end{equation}

\begin{figure}[h]
\begin{center}
\includegraphics[width=4.5in]{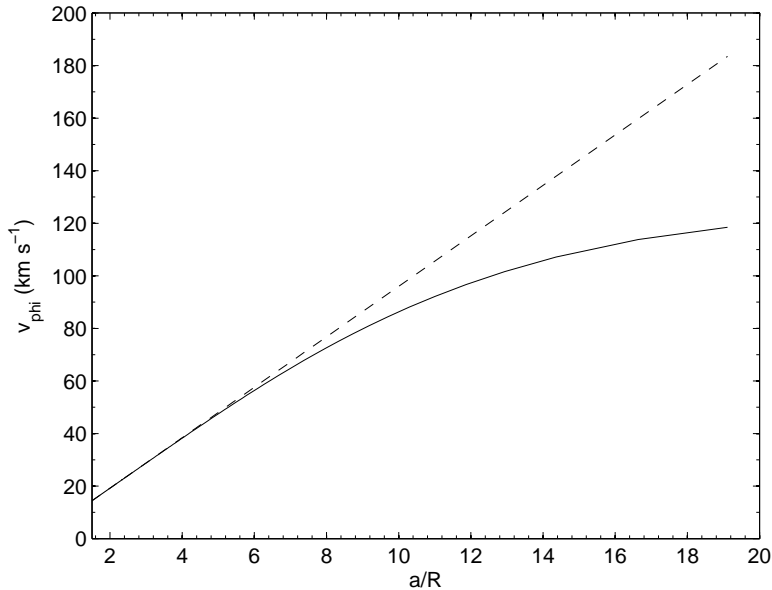}
\caption{Orbital velocity of equatorial plasma as a function of
distance from Saturn. Rigid corotation with Saturn's interior is
plotted as the dashed line; the magnetosphere lags the planetary
interior due to slowing of the ionosphere by magnetic torques. }
\label{fig:rot}
\end{center}
\end{figure}

Figure \ref{fig:rot} displays the run of $v_{\phi}=\Omega^M a$ vs
$a/R$ for our nominal parameters. The rotation curve is in reasonable
agreement with \emph{Voyager} measurements \citep{ric86}. At $a/R=4$,
corresponding to the orbit of Enceladus and the brightest portion of
the E-ring, $1-\Omega^M/\Omega\simeq 0.5\%$. By comparison, equation
\refnew{eq:DelOm} predicts the much smaller value, $3\times 10^{-5}$,
for the slippage of the rotation rate in the plasma tongue at $a=4R$
relative to that of the ionosphere.

\subsection{Why Is the Clock Imperfect?}

The ionospheric rotation rate responds to changes in a variety of
parameters, including the atmosphere's eddy diffusion coefficient, the
ionosphere's height-integrated conductivity, and the rate of mass loss
from the magnetosphere. All of these are likely to have a seasonal
dependence.  Solar weather, including short term variations in the
solar wind ram pressure and longer term variations over the solar
activity cycle also affect the conductivity in the auroral ionosphere.

\section{Summary}
\label{sec:sum}

Rotationally driven convection of magnetospheric plasma breaks the
axisymmetry of Saturn's external magnetic field. Field aligned
currents transfer angular momentum from the planet to a tongue of
outflowing plasma. This transfer slows the rate of rotation of the
ionosphere relative to that of the underlying atmosphere. The currents
are the source for the non-axisymmetric components of the field. The
common rotation rates of these components and Saturn's kilometric
radio (SKR) bursts is that of the plasma near the orbit of Enceladus, and
by extension the rotation rate in the ionosphere to which this plasma
is coupled. This rate tells us nothing about the rotation rate of
Saturn's deep interior. Of that we remain ignorant.

Magnetic perturbations with magnitudes similar to those observed by
\emph{Cassini} are produced for ${\dot M}\approx 10^4 \rm{~g~s}^{-1}$,
a value similar to estimates for the rate of production of plasma from
Saturn's E-ring.

Enhancement of the SKR occurs in a narrow range of longitudes where
the tip of the outgoing plasma stream connects to the auroral
ionosphere via field lines that are bowed outwards by currents
that supply the plasma's centripetal acceleration.

\end{document}